\documentclass[prl,aps,twocolumn,showpacs,floatfix]{revtex4}
\usepackage{graphicx,color}
\usepackage{dcolumn}
\usepackage{bm}
\usepackage{epstopdf}
\usepackage{color}

\newcommand{\beq}{\begin{eqnarray}}
\newcommand{\eeq}{\end{eqnarray}}

\newcommand{\hamil}{\mathcal{H}}
\newcommand{\timeve}{\mathcal{U}}

\newcommand{\dl}{_{\rm dl}}
\newcommand{\fl}{_{\rm fl}}

\pdfoutput=1

\begin{document}

\title{
Majorana Braiding Dynamics on Nanowires
}
\author{C\'assio Sozinho Amorim}
\affiliation{Department of Applied Physics, Nagoya University,
Nagoya 464-8603, Japan}
\author{Kazuto Ebihara}
\affiliation{Department of Applied Physics, Nagoya University,
Nagoya 464-8603, Japan}
\author{Ai Yamakage}
\affiliation{Department of Applied Physics, Nagoya University,
Nagoya 464-8603, Japan}
\author{Yukio Tanaka}
\affiliation{Department of Applied Physics, Nagoya University,
Nagoya 464-8603, Japan}
\author{Masatoshi Sato}
\email{msato@nuap.nagoya-u.ac.jp}
\affiliation{Department of Applied Physics, Nagoya University,
Nagoya 464-8603, Japan}
\date{\today}

\begin{abstract}
Superconductors hosting long-sought excitations called Majorana fermions
may be ultimately used as qubits of fault-tolerant topological quantum
computers. A crucial challenge toward the topological quantum computer
is to implement quantum operation of nearly degenerate quantum states as a
dynamical process of Majorana fermions. In this paper, we investigate
the braiding dynamics of Majorana fermions on superconducting
nanowires. In a finite size system, a non-adiabatic dynamical process dominates
the non-Abelian braiding that operates qubits of Majorana fermions. Our
simulations clarify how qubits behave in the real-time braiding
process, and elucidate the optimum condition of superconducting
nanowires for efficient topological quantum operation.  
\end{abstract}

\pacs{
}


\maketitle


{\it Introduction---}
Recent discovery of topological matters provides a novel
platform of quantum devices. 
In  particular , topological superconductors naturally realize yet-to-be
discovered excitations called Majorana fermions as a collective mode in
condensed matter physics \cite{TSN12, QZ11, Wilczek}.
Because of the self-antiparticle nature, the isolated Majorana zero modes
display unusual physical properties such as non-Abelian anyon
statistics, which is of extreme interest in realization of topological
quantum computer in reality.
%

Topological superconductivity was originally recognized in $p$-wave
spin-triplet superconductors \cite{RG00,Ivanov01,Kitaev01,Sato10}, however,  advance on our understanding
of topological matters enables us to design it even in a conventional
$s$-wave superconducting state \cite{Sato03, FK08, STF09, SLTS10}.
A recent proposed scheme to realize Majorana fermions by using the
spin-orbit interaction and Zeeman field \cite{SF09,STF09,STF10,SLTS10,Alicea10}
was eventually applied to a one-dimensional nanowire with proximity
induced $s$-wave pairing \cite{LSS10, ORO10}, 
which can be fabricated by the present experimental technique \cite{mourik12,deng12,das12,williams12,velhorst12,rokhinson12}. 
Furthermore,  varieties of proposals exist in order
to improve the experimental accessibility and controllability of
Majorana modes \cite{LTYSN10, SF10, Alicea12, ST13, Beenakker13, Hassler, STLSS10, KL12, SLHS12, CSSB13, SS12, NBTTN13, MPWB13}.

In topological quantum computation, quantum operations of qubits 
are implemented as an exchange process of Majorana zero modes. 
Thus, a crucial next step toward topological quantum computer
is to understand such an operation of collective excitations as a
time-dependent dynamical process.

In this paper, we investigate
the braiding dynamics of Majorana zero modes on superconducting
nanowires.
Generalizing proposed methods of Majorana braiding
\cite{AOROF11,SCT11,LWH12,KSS12,HOSRAO12,  STS10, LWL13, ZKM13, LNBY13, WRS13, KRO13,
HHFBAB13, KZB13, LL13, CVF14}, 
we consider a simpler cruciform junction of topologically non-trivial
superconducting nanowires.
This simple system functions as a quantum NOT gate of a Majorana qubit
by switching gates connecting the wires to the cross point.
Using this model, we simulate the Majorana braiding by solving the
time-dependent Bogoliubov de Genne equation for the nanowires.  
A non-adiabatic dynamical process dominates
the non-Abelian braiding that operates qubits of Majorana fermions. Our
simulations clarify how qubits behave in the real-time braiding
process, and elucidate the optimum condition of superconducting
nanowires for efficient topological quantum operation.  







{\it Majorana Braiding---}
In the low energy limit, 
one-dimensional
topological superconductors reduce
to a one-dimensional spinless $p_x$-wave superconductor. 
We adopt the spinless $p_x$-wave superconductor as a model to analyze
universal aspects of 
Majorana dynamics on nanowires,
\begin{eqnarray}
{\cal H}=-\mu\sum_{x=1}^{N}c^{\dagger}_xc_x-
\sum_{x=1}^{N-1}
\left(\lambda c_x^{\dagger}c_{x+1}+\Delta e^{i\theta}c_x c_{x+1}+{\rm
 h.c.}\right),
\nonumber\\ 
\label{TB_hamil}
\end{eqnarray}
where $c_x$ is a spinless fermion operator and $\mu$, $\lambda$ and
$\Delta e^{i\theta}$ are the chemical potential, the hopping integral, and the
$p$-wave pairing potential, respectively. ($\lambda>0$, $\Delta>0$.)
There are two different topological phases in the spinless $p_x$-wave
superconductor \cite{Kitaev01}.
When $|\mu|<\lambda$, the $p_x$-wave superconductor realizes a
topologically non-trivial superconducting state, and thus it supports a Majorana
fermion on each end. 
In contrast, when $|\mu|>\lambda$,  it becomes a topologically
trivial state without Majorana end modes. 
Below, we consider $p_x$-wave superconducting nanowires in the
topologically non-trivial phase. 

\begin{figure}[h]
\includegraphics[width=70mm]{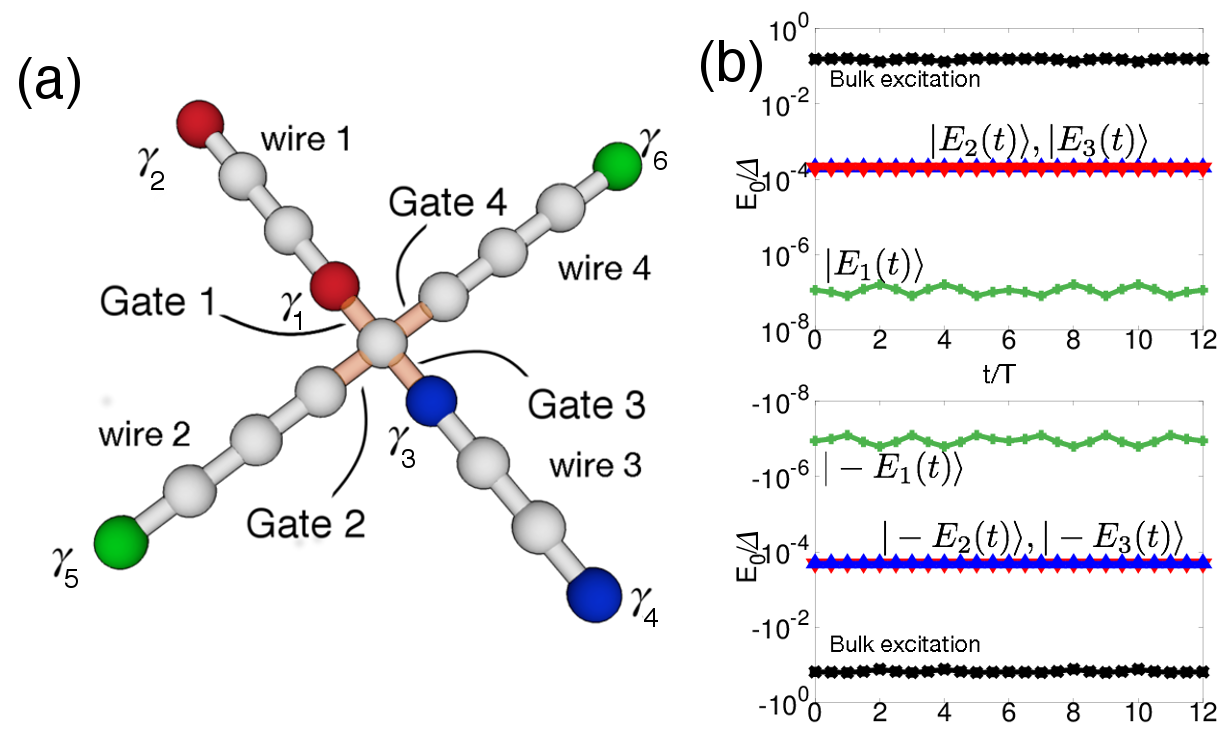}
\caption{ (color online). (a) Cross-shaped topological
 superconducting nanowire. The orange junctions connecting to the
 central site are cut/linked through gate potentials, effectively
 leaving 4 independently controlled wires. 
When gates 2 and 4 are
connected, while 1 and 3 are shut, six Majorana end modes  $\gamma_i$
 $(i = 1, 2, 3, 4, 5, 6)$ are obtained. 
(b) A typical energy
 spectrum of MBSs as a function of $t$, where the time-dependence is
 given by the gating process in Fig.\ref{fig:3}.
Here we take $\Delta=\lambda$, $\mu=0.7\lambda$ and $N=20$.
The color of the lines match the color of
 the modes in Fig.\ref{fig:2}(a). 
The finite coupling of MBS on the same wire slightly
lifts the zero-mode degeneracy.}
\label{fig:2}
\end{figure}

To braid the Majorana end states, we consider a
cruciform junction illustrated in Fig.\ref{fig:2} (a), where
four topologically non-trivial nanowires
(wire 1, 2, 3 and 4) are connected by four gates (gate 1, 2, 3 and 4). 
The hopping integral $\lambda$ and the paring potential
$\Delta$ at the gates are
tunable, so one can connect (disconnect) the wires by turning on (off)
these parameters at the gates.

\begin{figure}[t!]
\includegraphics[width=60mm]{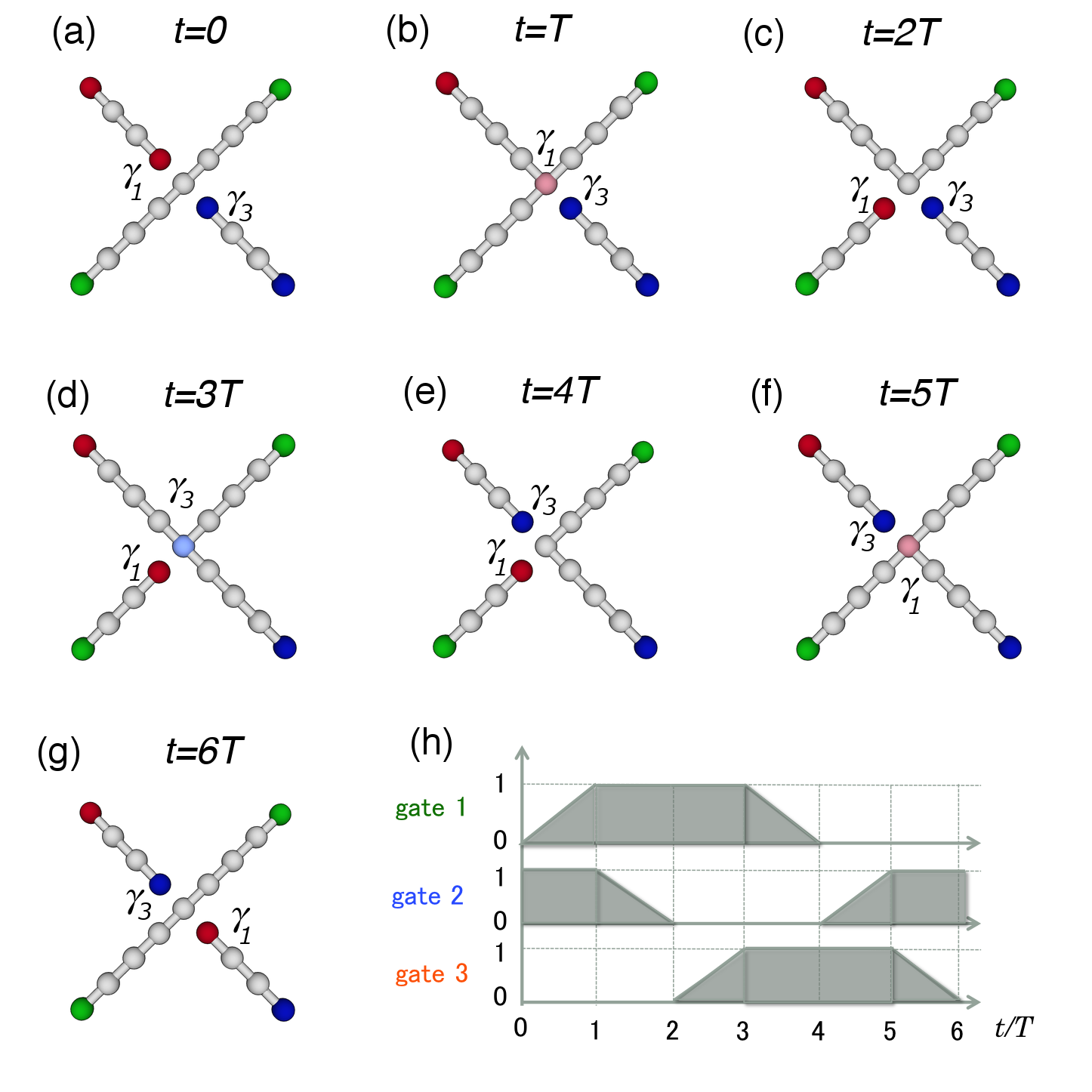}
\caption{(color online). The initial conditions of our system is shown in
 (a), which is operated according to dimensionless gate parameters shown
 in (h), being 0 a completely separated wire, and 1 a fully connected
 condition. $T$ is our gate operation time parameter. Majorana
 $\gamma_1$ is moved towards the center in (b) by connecting gate 1 and
 then to wire 2 in (c) by disconnecting gate 2.  Majorana $\gamma_3$ is
 moved  from wire 3 to 1 by a similar process  in (d) and (e), later
 taking $\gamma_1$ from wire 2 to 3, obtaining an interchange of
 $\gamma_1$ and $\gamma_3$.}
\label{fig:3}
\end{figure}

Now let us illustrate how one can exchange the Majorana end modes by switching
these gates of the cruciform junction.
Initially, we prepare the configuration of
Fig.\ref{fig:3}(a), where wires 2 and 4 are connected by turning on gates
2 and 4, while wires 1 and 3 are disconnected.
There are six Majorana end states in the
initial configuration since two Majorna end states at the inner edge of
wires 2 and 4 are gapped by the coupling at gates 2 and 4.

Counterclockwise exchange of the Majorana modes
$\gamma_1$ and $\gamma_3$, which are localized at the inner edges of
wires 1 and 3,  can be implemented as follows:
First, by turning on the gate 1 [Fig. \ref{fig:3}(b)] and then turning
off gate 2 [Fig. \ref{fig:3}(c)],
$\gamma_1$ moves to the inner edge of wire 2. 
Next, $\gamma_3$ moves to the inner edge of wire 1
by turning on gate 3 [Fig. \ref{fig:3}(d)] and then turning off gate 1
[Fig. \ref{fig:3}(e)].
Finally, $\gamma_1$ moves to the inner edge of wire 3 by turning on gate
2 [Fig. \ref{fig:3}(f)] and then turning off gate 3
[Fig. \ref{fig:3}(g)]. 
The final gate configuration is identical to the initial one, but 
$\gamma_1$ and $\gamma_3$ are exchanged.

In the initial configuration $(t=0)$, wire 1 and wire 3 are
isolated from others.
While wires 2 and 4 are connected to each other, 
they are also disconnected from the rest. 
Due to the finite length of wires, the mixing of Majorana modes occurs between $\gamma_1$ and $\gamma_2$, $\gamma_3$ and $\gamma_4$, and
$\gamma_5$ and $\gamma_6$, respectively \cite{Kitaev01,CLGS09, MM10, CLGS10, CGS11}. 
It induces 
the following effective coupling between zero modes, 
\begin{eqnarray}
{\cal H}_{\rm eff}=i \epsilon \gamma_1\gamma_2+i\epsilon\gamma_3\gamma_4+i\epsilon'
 \gamma_5\gamma_6,
\end{eqnarray}
where the constants $\epsilon$ and $\epsilon'$
are real because ${\cal H}$ should be hermitian, and 
$\epsilon$ is larger than $\epsilon'$ since the coupling between $\gamma_5$ and
$\gamma_6$ is weaker than the others.
Assuming the standard anti-commutation relation of Majorana zero modes,
{\it i.e.}
$\{\gamma_i,\gamma_j\}=2\delta_{ij}$, ${\cal H}$ can be recast into
\begin{eqnarray}
{\cal H}_{\rm eff}=\epsilon' \left(
2c_1^{\dagger}c_1-1
\right) 
+\epsilon \left(
2c_2^{\dagger}c_2 
+2c_3^{\dagger}c_3-2 
\right)
\end{eqnarray}
with the Dirac operators
\begin{eqnarray}
c_1=\frac{\gamma_5+i\gamma_6}{2}, 
\quad 
c_2=\frac{\gamma_3+i\gamma_4}{2},
\quad
c_3=\frac{\gamma_1+i\gamma_2}{2},
\end{eqnarray}
obeying $\{c_i, c_j^{\dagger}\}=\delta_{ij}$.
Therefore, the mixing results in three negative energy states
$|-E_i\rangle$ 
$(i=1,2,3)$ that are annihilated by $c_i$,
\begin{eqnarray}
c_i|-E_i\rangle=0 
\label{eq:negative_state}
\end{eqnarray}
and three positve partners $|E_i\rangle$  that are annihilated by
$c_i^{\dagger}$,  
\begin{eqnarray}
c_i^{\dagger}|E_i\rangle=0,
\label{eq:positive_state}
\end{eqnarray}
with $E_1=\epsilon'$ and $E_2=E_3=\epsilon$.
%

By switching the gates as described above, we can exchange the Majorana zero
modes $\gamma_1$ and $\gamma_3$.
A proper gating process for the non-Abelian braiding does not
merely interchange
$\gamma_1$ and $\gamma_3$, but also provide a
non-trivial relative phase between them,
$
\gamma_1\rightarrow -\gamma_3$,
$\gamma_3\rightarrow \gamma_1,
$
or 
$
\gamma_1\rightarrow \gamma_3
$,
$
\gamma_3\rightarrow -\gamma_1
$.
In both cases, if we exchange $\gamma_1$ and
$\gamma_3$ twice, $\gamma_1$ and $\gamma_3$ do not go back to the
original, but they acquire the minus sign,
$
\gamma_1\rightarrow -\gamma_1,
$
$
\gamma_3\rightarrow -\gamma_3. 
$
Therefore, after  exchange  $\gamma_1$ and $\gamma_3$ twice, 
the Dirac operators $c_2$ and $c_3$ transform into their conjugates
$-c_2^{\dagger}$ and $-c_3^{\dagger}$ as
\begin{eqnarray}
&&c_2=\frac{\gamma_3+i\gamma_4}{2}\rightarrow\frac{-\gamma_3+i\gamma_4}{2}
=-c_2^{\dagger}.
\nonumber\\ 
&&c_3=\frac{\gamma_1+i\gamma_2}{2}\rightarrow\frac{-\gamma_1+i\gamma_2}{2}
=-c_3^{\dagger},
\end{eqnarray}
The nontrivial transformation of $c_2$ and
$c_3$ implies that
the negative energy states 
$|-E_2\rangle$ and $|-E_3\rangle$, which are annihilated by $c_2$
and $c_3$, respectively, end up as the positive energy partners
$|E_2\rangle$ and $|E_3\rangle$, and vice versa, after the exchange process. 
In other words, if we choose these negative energy states as an initial state,
the final state is orthogonal to the initial one. 
This complete interference is a direct signal of the non-Abelian anyon
statistics: Indeed if the system obeys an ordinary Abelian
statistics, any exchange process results in a phase factor for any
initial state, so the final state cannot be orthogonal to the initial one.
The above exchange process defines a quantum NOT gate for
Majorana qubits $(|E_2\rangle, |-E_2\rangle)$ and $(|E_3\rangle,
|-E_3\rangle)$.


Whereas the above procedure eventually works well as is shown below, 
the actual implementation needs a careful consideration for the gating.
In Fig.\ref{fig:2} (b), we show lower energy eigenvalues of
the system as a function of $t$. 
The eigen energies $E_1(t)$, $E_2(t)$, $E_3(t)$ and their negative energy
partners correspond to six Majorana zero modes of the system, where 
$E_2(t)$ and $E_3(t)$ are degenerate within numerical
accuracy as well as their negative energy partners are.
At $t=0$, these eigen energies coincide with $E_i$ $(i=1,2,3)$ in the
above,
$
E_i(0)=E_i,
$
and we also have $E_i(6T)=E_i(12T)=E_i$ since 
the system goes back to the initial
configuration at $t=6T$ and $12T$.
We note here that there is no level crossing in
the energy spectrum in Fig.\ref{fig:2} (b), as expected from the von
Neumann-Wigner theorem \cite{NW1929}.
Therefore, a non-adiabatic transition is needed to achieve the
non-Abelian braiding discussed in the above, since
any state cannot be different from the original under an adiabatic process. 
Namely, the gating process in
Fig.\ref{fig:3} should not be too slow.
The non-adiabatic transition is not a classical Landau-Zener
transition, because the level spacing rarely depends on $t$ and there is no level
approaching to each other at a particular time.
We can also argue that a proper gating process should not be too fast at
the same time. 
A fast gating process may create bulk
excitations on nanowires, which may give rise to problematic decoherence
of Majorana qubits.
Therefore, the gating process for non-Abelian braiding should be performed at a
proper range of speed.  

Below we operate the gates 1, 2, 3 in accordance with the time-sequence
diagram in Fig.\ref{fig:3} (h).
The gating speed can be controlled by an adiabatic parameter $T$: 
The gate operation becomes slower (faster) and more adiabatic
(non-adiabatic) for larger (smaller) $T$.
A moderate $T$ is required to realize the non-Abelian braiding.

{\it Braiding Dynamics---}
We now numerically simulate the Majorana braiding process in Fig.\ref{fig:3}.  
To numerically evaluate our system, we take each wire length to be the
same 
with one central site linking them.
Each gate is represented as a factor $g_i\in[0,1]$ ($i=1,2,3,4$)
multiplying the link on the gates in real space \footnote{For details, see
Supplementary Material.}.
The dynamics of the system is described by the time-dependent
Bogoliubov-de Genne equation
\begin{eqnarray}
i\hbar\frac{\partial}{\partial t}\Psi(t)={\cal H}(t)\Psi(t), 
\end{eqnarray}
where $\Psi(t)$ is the quasiparticle wavefunction in the Nambu representation.
The evaluation of the wavefunction during a time ${\mathit \Delta} t$ is
given by
$\Psi(t+{\mathit \Delta} t)={\cal U}(t+\mathit{\Delta}t; t)\Psi(t)$ with the
time-evolution operator ${\cal U}(t+{\mathit \Delta} t;t)$,
\begin{eqnarray}
{\cal U}(t+\mathit{\Delta} t,t)=T\exp\left[-i\int^{t+\mathit{\Delta} t}_{t} d\tau {\cal
			     H}(\tau)\right],  
\end{eqnarray}
which is well-approximated as
$
{\cal U}(t+\mathit{\Delta} t,t)\approx\exp\left[-i
{\cal H}(t)\mathit{\Delta} t\right]
$,   
within numerical errors for a sufficiently short $\mathit{\Delta} t$.   
To achieve a correct wavefunction change in time, we further expand the
time-evolution operator in terms of Chebishev polynomials \cite{TK84,LWH12},
which can be retrieved recursively, that is
\begin{eqnarray}
\mathcal{U}(t+\mathit{\Delta} t;t)
&=&\exp\left[ -i\frac{\hamil(t)}{E_0}\mathit{\Delta} tE_0\right]
= \exp\left[ -i\tilde{H}(t)\mathit{\Delta} \tau\right]\nonumber\\
&=& \sum_{k=0}^{\infty}c_k(\mathit{\Delta}\tau)T_k(\tilde{H}(t)),
\label{eq:Chebisev}
\end{eqnarray}
where $E_0\equiv \max|\langle\Psi|\hamil|\Psi\rangle|$ normalizes the
Hamiltonian to avoid singularities of Chebishev polynomials and
\begin{eqnarray}
&&c_k(\mathit{\Delta}\tau)
= \left\{ \begin{array}{ll}
J_0(\mathit{\Delta}\tau)&(k=0)\\
2(-i)^kJ_k(\mathit{\Delta}\tau)& (k\ge1)
\end{array}\right.\\[5pt]
&&T_0=1, \quad 
T_1(\tilde{H})=\tilde{H},\nonumber\\
&&T_{k+1}(\tilde{H}) = 2\tilde{H}T_k(\tilde{H})-T_{k-1}(\tilde{H}),
\end{eqnarray}
constitute our expansion terms. 
Here $\tilde{H}={\cal H}/E_0$,  $\mathit{\Delta}\tau=\mathit{\Delta} t E_0$, and $J_k$ are the Bessel functions of first
kind.
For small $\mathit{\Delta} \tau$, 
the coefficients $c_k(\mathit{\Delta}\tau)$ rapidly converge to zero as $k$
increases.
Thus keeping the first few expansion terms in the
right hand side of Eq.(\ref{eq:Chebisev}) is enough to reach numerically
reliable results.

\begin{figure}[h]
\includegraphics[width=85mm]{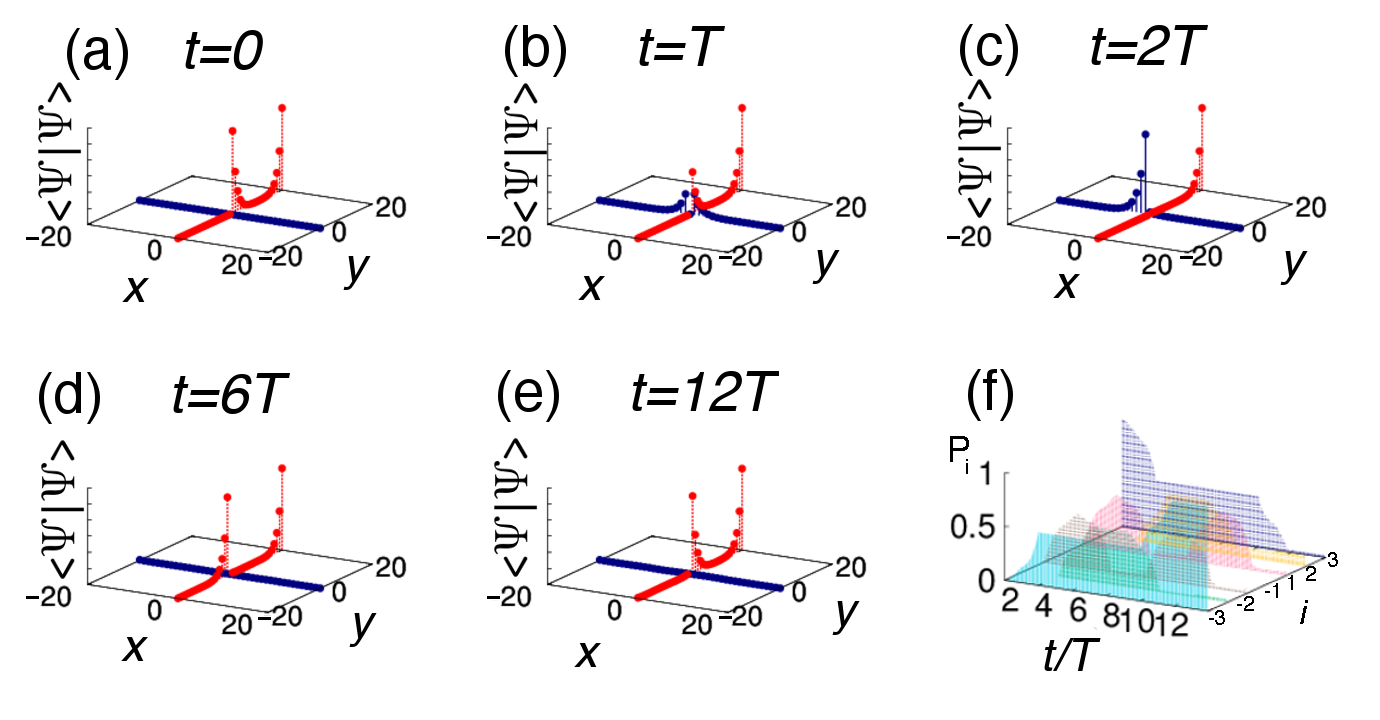}
\caption{(color online). $\gamma_2$ braiding with $T = 100/\Delta$.
We take $\Delta=\lambda$, $\mu=0.7\lambda$ and $N=20$.  We start with
 the squared wavefunction in (a) localized in wire 1, which can be seen
 to migrate towards the center only for $\gamma_1$ in (b). The complete
 transfer of $\gamma_1$ to wires 2 and 3 are respectively seen in (c) and
 (d). Final state at (e) indicates the return of the MBS to wire 1. (f)
Projection of the wave function on the instantaneous
 eigenstates, $P_{\pm i}(t)=\langle \Psi(t)|\pm E_i(t)\rangle$
 ($i=1,2,3$).
We can observe our wavefunction begins completely on $|E_3\rangle$ and after
 superposing on other eigenstates, completely transfers to $|-E_3\rangle$.
} 
\label{fig:5}
\end{figure}

Figure \ref{fig:5} is one of the main results in this paper.
In Fig.\ref{fig:5}, we illustrate how the wavefunction 
evolves in time in our numerical simulation of the non-Abelian braiding.
We choose $|E_3\rangle$ as the initial state at $t=0$, and take $T=100/\Delta$.
It demonstrates that only the inner part of the wave function moves in time,
which exactly corresponds to the movement of Majorana mode $\gamma_1$. 
At $t=6T$, although the gate configuration goes back to the initial one,  
the inner part of the wave function moves to the inner edge of wire
3, which indicates that $\gamma_1$ is successfully interchanged with $\gamma_3$.
Then finally, the inner part goes back to the initial position at $t = t_{\rm f} \equiv 12T$.

We project the same wavefunction into the instantaneous
eigenstates $|\pm E_i(t)\rangle$ $(i=1,2,3)$ in Fig.\ref{fig:5} (f). 
Initially, the wavefunction consists of only $|E_3\rangle$, but 
after the gating process starts, 
the wave function quickly spreads over the eigenstates $|-E_3\rangle$
and
$|\pm E_3\rangle$.
Nevertheless, the final state ends up at $|-E_3\rangle$, as expected
as the non-Abelian braiding process mentioned above.

\begin{figure}[h]
\includegraphics[width=80mm]{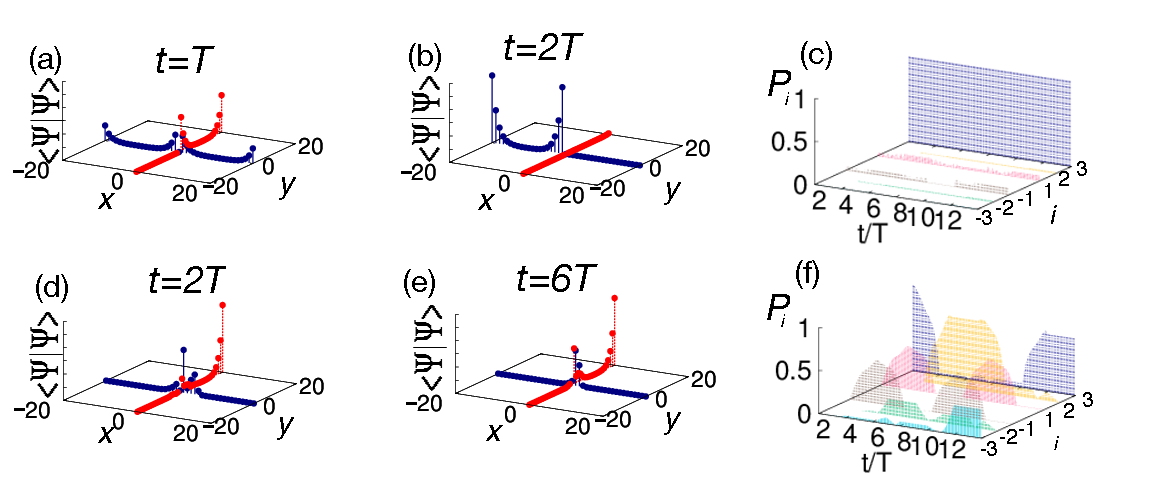}
\caption{(color online). Behavior outside braiding (good)
 conditions. (a) and (b) represent the evolution of the system on
 adiabatic limit ($T=100000/\Delta$), where both Majorana edge-states
 migrate to the next wire, as can be seen from (b), consequently
 remaining on the same state, as can be observed from the projection of
 the wavefunction on the instantaneous eigenstates in (c). (d) and (e)
 account for the opposite, fast regime ($T=1/\Delta$), with bulk excitations
 visible in (d) and a more erratic spread over the eigenstates in (f).
We take $\Delta=\lambda$, $\mu=0.7\lambda$, and $N=20$.} 
\label{fig:5.5}
\end{figure}

If one operates the gates too slowly or too quickly, the
non-Abelian braiding fails.
For example, in the slow gating with $T=100000/\Delta$, both
inner and outer Majorana modes of wire 1 move together in time, in which the
state mostly stays at
the instantaneous eigenstate $|E_3(t)\rangle$, as expected by the
adiabatic theorem. 
On the other hand, for the quick gating with $T=1/\Delta$,
the wave function extends over all eigenstates $|\pm E_i(t)\rangle$
$(i=1,2,3)$, and it never goes back to the initial state.
In the latter case, the wave function in the final state also spreads
over nanowires in space, which suggests that bulk modes are excited
during the gating.
We exemplify these unsuccessful braiding in Fig. \ref{fig:5.5}.

\begin{figure}[h]
\includegraphics[width=70mm]{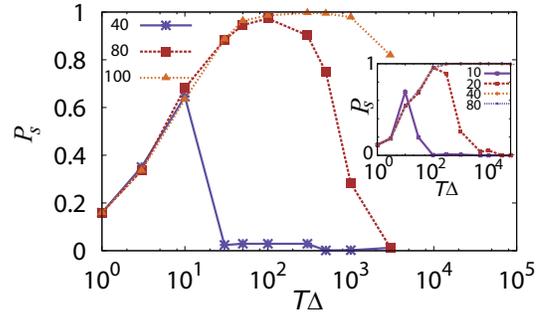}
\caption{(color online). Success rate $P_{\rm s}$ of the Majorana braiding.
Each line accounts for a different length (in site number) of each wire (1-4), with $x$-axis being our operation time $T$, and $y$-axis the success rate $P_s$. 
We take $\Delta=0.1\lambda$ and $\mu=0.7\lambda$. The inset shows
 $P_{\rm s}$ for $\Delta=\lambda$.
The red lines (80 and 20 sites/wire) draw a clear separation between its adiabatic (slow, large $T$) and dissipative (fast, small $T$) domains, suggesting an appreciable necessity for both size and time scale adjustment.}
\label{fig:6}
\end{figure}

To quantify the non-Abelian braiding, 
we introduce its success rate $P_{\rm s}$ as the
probability that the final state $|\Psi(t_{\rm f})\rangle$ is found to be
the desired state $|-E_3\rangle$,
\begin{eqnarray}
P_{\rm s}=|\langle \Psi(t_{\rm f})|-E_3\rangle|^2 
\end{eqnarray}
In Fig.\ref{fig:6}, we plot the success rate $P_{\rm s}$ versus the
adiabatic parameter $T$, with various
wire lengths.
The data indicates that a longer wire is desirable for the
non-Abelian braiding.
For longer wires, the success rate can reach the maximum, {\it i.e.}
$P_{\rm s}=1$ for large $T$.
In a shorter wire, on the other hand, a Majorana end mode is fairly
coupled with the Majorana mode on the other end, so they tend to move
together, resulting in an adiabatic process even for a moderate $T$.   
We also find that 
a quicker gating fails to achieve the non-Abelian
braiding for any length of wires, since it excites undesirable bulk modes.

Finally, from our numerical results, we evaluate the optimal condition
for non-Abelian braiding. 
We note that $T$ should be larger than the inverse of the bulk gap
$1/\Delta$, not to excite bulk modes.
Our numerical data determines how large it should be. 
Figure \ref{fig:6} indicates that 
the lower bound $T$ is not a merely $O(1/\Delta)$, but it is
evaluated as $T_{\rm min}\sim O(10^2/\Delta)$.
For a typical superconducting state with $\Delta=O(1)$ K,
$T_{\rm min}$ can be a few nanoseconds. 
On the other hand, the upper bound of $T$ can be determined as follows.
As we illustrated in the above, the non-Abelian braiding is realized as an
non-adiabatic process between Majorana modes. 
Thus, for Majorana modes with energy $\epsilon$, 
$\epsilon T$ should not be too large.
Our numerical results imply that the success rate of the non-Abelian
braiding $P_{\rm s}$ reaches almost the maximum when $\epsilon T$ is less than
$O(10)$.
The latter condition can be easily met for long wires, since $\epsilon$
scales as $\epsilon \sim e^{-N/l_0}$. 
It is also found in Fig. \ref{fig:6} that the quantum limit of
superconducting state,
i.e. $\Delta/\lambda=1$, requires less numbers of sites
for the non-Abelian braiding, which is preferable if one realizes
topological supercondicting wires as a chain of quantum dots.\cite{SS12}
The authors are grateful to J. D. Sau and K. T. Law
for fruitful discussions. This work was supported by the JSPS
(No.25287085) and KAKENHI
Grants-in-Aid (No.22103005) from MEXT.
C.S.A. is supported by
MEXT Scholarship (kokuhi-gaikokujin-ryugakusei 2013) 

\bibliography{braiding}

\newpage

\appendix

\setcounter{equation}{0}
\centerline{\large \bf Supplementary Material}

\section{Calculation Method}
The Hamiltonian in Eq.~(\ref{TB_hamil}) can be recast under
Bogoliubov-de Gennes representation as
\begin{eqnarray}
\mathcal{H}^{\mathrm{BdG}}&=&-\left[\frac{\mu}{2}
+\frac{\lambda}{2}\left( \cos k_x+\cos k_y\right)\right]\tau_z\nonumber\\
&&-\frac{\Delta}{2}e^{i\theta}\left(\sin k_x\tau_y + \sin k_y\tau_x\right),
\end{eqnarray}
which in turn can be rewritten in real space in matrix form, defining our Hamiltonian as $H=\sum_{ij}c^{\dagger}_i\mathcal{H}_{{\bm i},{\bm j}}c_j$ on Nambu basis,
\begin{eqnarray}
&&\mathcal{H}_{{\bm i},{\bm i}}=
\left( \begin{array}{cc}
-{\mu} & 0 \\
0 & {\mu}
\end{array}
\right),
\nonumber\\
&&\mathcal{H}_{{\bm i}\pm \hat{\bm e}_x,{\bm i}}=
\left( \begin{array}{cc}
-\lambda/2 & \mp\Delta e^{i\theta}/2 \\
\pm\Delta e^{-i\theta}/2 & \lambda/2
\end{array}\right),
\nonumber\\
&&\mathcal{H}_{{\bm i}\pm\hat{\bm e}_y, {\bm i}}=
\left( \begin{array}{cc}
-\lambda/2 & \mp\Delta e^{i\theta}/2 \\
\mp\Delta e^{-i\theta}/2 & \lambda/2
\end{array}
\right),
\end{eqnarray}
where ${\bm i}$ represents site position, assuming lattice constant to be 1.
To numerically evaluate our system, we take each wire length to be $n$
sites long with one central site linking them.
Each gate is represented as a factor $g_i\in[0,1]$ ($i=1,2,3,4$)
multiplying the Hamiltonian elements between the central site and its
neighbors in real space, 
\begin{eqnarray}
&&{\cal H}_{\hat{\bm e}_y, 0}=g_1
\left(
\begin{array}{cc}
-\lambda/2 & -ie^{i\theta}\Delta/2\\
 -ie^{-i\theta}\Delta/2 & \lambda/2
\end{array}
\right),
\nonumber\\
&&{\cal H}_{-\hat{\bm e}_x, 0}=g_2
\left(
\begin{array}{cc}
-\lambda/2 & e^{i\theta}\Delta/2\\
 -e^{-i\theta}\Delta/2 & \lambda/2
\end{array}
\right),
\nonumber\\ 
&&{\cal H}_{-\hat{\bm e}_y, 0}=g_3
\left(
\begin{array}{cc}
-\lambda/2 & ie^{-i\theta}\Delta/2\\
 ie^{-i\theta}\Delta/2 & \lambda/2
\end{array}
\right),
\nonumber\\
&&{\cal H}_{\hat{\bm e}_x, 0}=g_4
\left(
\begin{array}{cc}
-\lambda/2 & -e^{i\theta}\Delta/2\\
 e^{-i\theta}\Delta/2 & \lambda/2
\end{array}
\right).
\end{eqnarray}
For a linear gate operation
in the interval of time $T$ as described in this work, a gate being turned on (off)
is taken to evolve as $g_i=t/T$ ($g_i=1-t/T$), counting the time $t$ from the beginning of the operation.
More general functions may be used for operating both gates smoothly at
the same time.
Finally, to achieve the wavefunction change in time, we expand the
time-evolution operator in terms of Chebishev polynomials\cite{TK84},
 which can be retrieved recursively, that is
\begin{eqnarray}
\mathcal{U}(t+\mathit{\Delta} t;t)
&=&\exp\left[ -i\frac{\hamil(t)}{E_0}\mathit{\Delta} tE_0\right]
= \exp\left[ -i\tilde{H}(t)\mathit{\Delta} \tau\right]\nonumber\\
&=& \sum_{k=0}^{\infty}c_k(\mathit{\Delta}\tau)T_k(\tilde{H}(t)),
\end{eqnarray}
 $E_0\equiv \max|\langle\Psi|\hamil|\Psi\rangle|$ normalizes the
Hamiltonian to avoid singularities and
\begin{eqnarray}
&&c_k(\mathit{\Delta}\tau)
= \left\{ \begin{array}{ll}
J_0(\mathit{\Delta}\tau)&(k=0)\\
2(-i)^kJ_k(\mathit{\Delta}\tau)& (k\ge1)
\end{array}\right.\\[5pt]
&&T_0=1, \quad 
T_1(\tilde{H})=\tilde{H},\nonumber\\
&&T_{k+1}(\tilde{H}) = 2\tilde{H}T_k(\tilde{H})-T_{k-1}(\tilde{H}),
\end{eqnarray}
constitute our expansion terms. $J_k$ are the Bessel functions of first
kind, and their value is used to choose truncation point under double
precision. While one can apply the above operator successively,
obtaining the desired wavefunction in time $t$, this process is
needlessly slow if done completely in double precision. Mixed precision\cite{GST07,BBDKLLLT09,GGW13}
can be efficiently applied for a boost in calculation speed if instead
of calculating the wavefunction, only its variation is evaluated. 
Explicitly, this can be illustrated on the following steps: 
\begin{eqnarray}
&&\Psi(t+\mathit{\Delta}t)\dl=\timeve(t+\mathit{\Delta}t;t)\Psi(t)\dl\nonumber\\
&&\rightarrow \mathit{\Delta}\Psi\dl=(\timeve(t+\mathit{\Delta}t;t)-1)\Psi(t)\dl\label{stddl}\\
&&\mathit{\Delta}\Psi=_{\rm dl}
\left(\sum_{k=0}^{k'}c_kT_k(\tilde{H})-1\right)\fl\Psi(t)\fl
\nonumber\\
&&=_{\rm
 dl}\left[(c_0-1)_{\rm fl}+\sum_{k=1}^{k'}c_{k,{\rm fl}}
T_k(\tilde{H})_{\rm fl}\right]\Psi(t)_{\rm
 fl}
\nonumber\\
&&\Psi(t+\mathit{\Delta} t)=_{\rm dl}\Psi(t)_{\rm dl}+\mathit{\Delta}\Psi_{\rm dl}
\end{eqnarray}
Here the subscript ${\rm dl}$ (${\rm fl}$) represents a double (single)
precision conversion/variable. For example, equation \ref{stddl} represents
a common time-evolution process done with double precision variables.
The following expression claims a conversion of $\Psi(t)$ from double to single precision,
as well as a single precision expansion of the time-evolution operator, while $=\dl$
implies that these data should be converted to double precision when adding to build up $\mathit{\Delta}\Psi$.
Similarly, on the following line we point that the zero-order term should be 
changed to single precision after calculating it in double precision, and each other expansion term is calculated with single precision $T_k$ and $c_k$. Shortly, each term in our expansion is
calculated as a vector in single precision, being later added as  a
double precision variation to the double precision wavefunction. Note
that the wavefunction is converted to single precision for time
evolution, but computed as double precision in the end of each step. In
fact, we start our evaluation with an eigenstate of the Hamiltonian
taken in double precision, and only its variation, which corresponds to
most of the computations, is found in single precision steps. 

It is important to note that this method relies on the constraint of small $dt$
to work, as well as small and smooth wavefunction variation, otherwise single precision
computation of the expansion terms may cumulate a large error. Nevertheless, this constraint
is also required for the very numeric expansion of the time-evolution operator, in order to ignore its
intrinsic time-ordering operator to a good approximation. In other words, the possibility to
evaluate time-evolution of the wavefunction in real space-time already gives us the possibility
of a mixed precision method. Concretely, in our case each 
$\mathit{\Delta}\Psi\sim10^{-7}$, which in single precision
 allows for good enough numeric results in the $10^{-7}\sim10^{-14}$ range,
  which accounts for first order terms, as well as $10^{-9}\sim10^{-16}$ 
  for the next order, all of them well fit in the limit of double precision, up to $10^{-15}$.
  Therefore, summing these terms up on double precision avoids 
  greater errors from ignoring their contribution, 
  which forcedly would happen if they were added completely in single precision.

\end{document}